# Three-Dimensional Topological Semimetal/Insulator States in $\alpha$-Type Organic Conductors with Interlayer Spin-Orbit Interaction

Toshihito Osada*

*Institute for Solid State Physics, University of Tokyo,*

*5-1-5 Kashiwanoha, Kashiwa, Chiba 277-8581, Japan.*

We have studied the tight-binding model for the $\alpha$-type layered organic conductors, $\alpha$-(ET)$_2$I$_3$ and $\alpha$-(BETS)$_2$I$_3$, with a uniform interlayer coupling accompanied by spin-orbit interaction originating from the I$_3^-$ anion potential. The model preserves the time reversal and inversion symmetries. In $\alpha$-(ET)$_2$I$_3$, the interlayer spin-orbit coupling realizes the experimentally suggested Dirac semimetal state with inversion symmetry. In contrast, the inversion breaking in interlayer hoppings realizes the Weyl semimetal state without spin-orbit coupling. In $\alpha$-(BETS)$_2$I$_3$, the proposed strong topological insulator is hardly realized with inversion symmetry.

The electronic state of the layered organic conductors $\alpha$-(ET)$_2$I$_3$ (an abbreviation for $\alpha$-(BEDT-TTF)$_2$I$_3$) and $\alpha$-(BETS)$_2$I$_3$, in which ET or BETS conducting layers and I$_3^-$ anion layers are alternately stacked, has been treated as a two-dimensional (2D) Dirac fermion (DF) system [1-11]. ET or BETS molecules form an anisotropic triangular lattice on the conducting layers, and I$_3^-$ anions form a rhombic lattice between the conducting layers [2]. In this $\alpha$-type configuration, the unit cell contains four molecules and two anions (Fig. 1(a)). Four $\pi$-bands are constructed from the HOMOs of the four molecules. Since the interlayer coupling, i.e. interlayer hopping across an anion layer is sufficiently weak, the electron systems are usually considered to be 2D systems. In 2006, it was theoretically suggested that $\alpha$-(ET)$_2$I$_3$ is the second DF system after graphene [3, 4]. In a suitable parameter range [5], the third and fourth bands touch at two nodal points in the 2D Brillouin zone (BZ) forming two tilted Dirac cones, and the Fermi level is stoichiometrically located at the Dirac point. This 2D DF state is topologically protected by the generalized chiral symmetry [6]. In real $\alpha$-type organic conductors, the gapless (massless) DF state is realized in $\alpha$-(ET)$_2$I$_3$ under high pressures suppressing the charge order, and the gapped (massive) DF state is realized in $\alpha$-(BETS)$_2$I$_3$ even at ambient pressure. The tight-binding calculations using the extended Huckel molecular orbitals and the first-principles calculations have confirmed the appearance of the 2D DF state in $\alpha$-(ET)$_2$I$_3$ [3, 7] and $\alpha$-(BETS)$_2$I$_3$ [8, 9]. Experimentally, the 2D DF states have been directly confirmed by the $\pi$-Berry phase of Shubnikov-de Haas oscillations using $\alpha$-(ET)$_2$I$_3$ and $\alpha$-(BETS)$_2$I$_3$ samples doped with contact charge [10, 11].

The spin-orbit interaction has been considered to be negligible in layered organic conductors consisting of light atoms. However, its importance has been pointed out based on the first-principles calculations, and the spin-orbit coupling (SOC) strength in the conducting layer has been estimated to be 1 ~ 2 and 5 ~ 10 meV for ET and BETS layers, respectively [12]. Based on this, possible 2D topological insulator (TI) state has been discussed in the α-type organic 2D DF systems with in-plane SOC [13]. This is an organic analogue of the Kane-Mele model for graphene [14]. At ambient pressure, α-(BETS)$_2$I$_3$ undergoes a metal-insulator crossover at around 50K. This is considered to be the temperature-induced dimensional crossover in the 2D TI state (2D massive DF state) with a small SOC gap due to the rather large in-plane SOC. The 2D TI state (3D weak TI state in bulk) has been predicted by first-principles calculations [8] and suggested experimentally by the anomalous transport suggesting surface states [15].

Recently, three-dimensional (3D) topological properties have been reported at low temperatures where the effect of interlayer coupling, i.e. interlayer hopping via an anion layer becomes significant [15, 20, 21]. When uniform interlayer coupling is simply introduced into the 2D massless DF system in α-(ET)$_2$I$_3$, it is expected to become a 3D semimetal with two straight nodal lines as will be seen later. However, the 3D nodal-point semimetal state has been theoretically predicted as a many-body topological phase considering electron correlation and multiple interlayer transfers [16]. In addition, the chiral-anomaly-related transport phenomena specific to the Weyl/Dirac nodal-point semimetals have been discussed [17]. This theory seems to be an extended version of the topological Mott insulator [18, 19]. It has been discussed that time reversal symmetry (TRS) and space inversion symmetry (SIS) are broken in this state. Experimentally, the coherence peak of the interlayer magnetoresistance, indicating the existence of a 3D Fermi surface, has been observed in α-(ET)$_2$I$_3$ [20]. In addition, negative longitudinal magnetoresistance and planar Hall effect, suggesting the chiral anomaly of nodal points, have been observed in α-(ET)$_2$I$_3$ at low temperatures [21]. On the other hand, when the interlayer coupling is not negligible in the 2D TI state of α-(BETS)$_2$I$_3$, the system is expected to become a 3D weak TI with surface states only on the side surfaces [8]. However, the surface transport over the whole surface, which is specific for 3D strong TIs, has been experimentally observed in α-(BETS)$_2$I$_3$ at low temperatures [15].

In this paper, we discuss the possible topological state in α-type organic conductors α-(ET)$_2$I$_3$ and α-(BETS)$_2$I$_3$ at low temperatures where interlayer coupling, i.e. interlayer hopping via the I$_3^-$ anion layer becomes non-negligible. We consider uniform interlayer coupling accompanied by SOC without any TRS and SIS breaking. The interlayer SOC originates mainly from the I$_3^-$ anion potential. It should be emphasized that even if the interlayer transfer integral is small, the influence of this interlayer SOC cannot be ignored when considering interlayer hopping, as electrons pass through the I$_3^-$ anion potential.

The schematic crystal lattice of α-type organic conductors is shown in Fig. 1(a). We extend the 2D tight-binding model given in Ref. [13] to the 3D model with simple interlayer coupling between the same molecular sites on the neighboring layers as shown in Fig. 1(a). In this model, we introduce an interlayer SOC (strength $\lambda$') associated with interlayer hopping in addition to the in-plane SOC (strength $\lambda$). Reflecting the configuration of I$_3^-$ anions at the top and bottom of the 2D layer, the in-plane SOC adds the imaginary contribution $\pm i\lambda b_i$ to the in-plane transfer integrals $b_i$ ($i$=1, 2, 3, and 4) between A-A' and B-C chains [13]. Since the interlayer hopping paths penetrate the I$_3^-$ anion layer, the interlayer SOC caused by the I$_3^-$ anion potential is considered to have non-negligible effects. The $I_3^-$ anion configuration is asymmetric around the interlayer hopping paths at sites A and A', so that a finite electric field **E** exists in the *y*-direction. The hopping electron (**p** // *z*-axis) feels the effective magnetic field ($\propto \mathbf{p} \times \mathbf{E}$) in the *x*-direction, resulting in the additional SOC contribution $i\lambda' t_A \sigma_x$ and $-i\lambda' t_{A'} \sigma_x$ with the Pauli matrix $\sigma_x$. Here, $t_A$ ($t_{A'}$) is the interlayer transfer integral between A (A') sites in neighboring layers. In contrast, no SOC contribution appears in the interlayer hopping at sites B and C, whose interlayer transfer integrals are $t_B$ and $t_C$ respectively, because the anion configuration is symmetric.

The tight-binding Hamiltonian is represented by the following 8×8 matrix $H(\mathbf{k})$, whose bases are the Bloch sums constructed from the HOMOs of the molecular sites A, A', B, and C with spin $\sigma_z = \pm 1$.

$$H(\mathbf{k}) = \begin{pmatrix} M(\mathbf{k}) & \Gamma(\mathbf{k}) \\ \Gamma(\mathbf{k})^\dagger & M(-\mathbf{k})^* \end{pmatrix}. \tag{1}$$

Here, the 4×4 matrices $M(\mathbf{k})$ and $M(-\mathbf{k})^*$ represent the up-spin ($\sigma_z = +1$) and down-spin ($\sigma_z = -1$) systems, respectively, and $\Gamma(\mathbf{k})$ represents the mixture of the two spin systems;

$$M(\mathbf{k}) = \begin{pmatrix} t_A(e^{-i\mathbf{c}\cdot\mathbf{k}} + e^{i\mathbf{c}\cdot\mathbf{k}}) & M_{AA'}(\mathbf{k}) & M_{AB}(\mathbf{k}) & M_{AC}(\mathbf{k}) \\ M_{AA'}(\mathbf{k})^* & t_{A'}(e^{-i\mathbf{c}\cdot\mathbf{k}} + e^{i\mathbf{c}\cdot\mathbf{k}}) & M_{A'B}(\mathbf{k}) & M_{A'C}(\mathbf{k}) \\ M_{AB}(\mathbf{k})^* & M_{A'B}(\mathbf{k})^* & t_B(e^{-i\mathbf{c}\cdot\mathbf{k}} + e^{i\mathbf{c}\cdot\mathbf{k}}) & M_{BC}(\mathbf{k}) \\ M_{AC}(\mathbf{k})^* & M_{A'C}(\mathbf{k})^* & M_{BC}(\mathbf{k})^* & t_C(e^{-i\mathbf{c}\cdot\mathbf{k}} + e^{i\mathbf{c}\cdot\mathbf{k}}) \end{pmatrix},$$

$$\Gamma(\mathbf{k}) = \begin{pmatrix} i\lambda' t_A(e^{-i\mathbf{c}\cdot\mathbf{k}} - e^{i\mathbf{c}\cdot\mathbf{k}}) & 0 & 0 & 0 \\ 0 & -i\lambda' t_{A'}(e^{-i\mathbf{c}\cdot\mathbf{k}} - e^{i\mathbf{c}\cdot\mathbf{k}}) & 0 & 0 \\ 0 & 0 & 0 & 0 \\ 0 & 0 & 0 & 0 \end{pmatrix}.$$

The matrix elements are given by $M_{AA'}(\mathbf{k}) = a_2 e^{i\mathbf{k}\cdot\boldsymbol{\tau}_1} + a_3 e^{-i\mathbf{k}\cdot\boldsymbol{\tau}_1}$, $M_{BC}(\mathbf{k}) = a_1 e^{i\mathbf{k}\cdot\boldsymbol{\tau}_1} + a_1 e^{-i\mathbf{k}\cdot\boldsymbol{\tau}_1}$,

$M_{AB}(\mathbf{k}) = b_2(1+i\lambda)e^{i\mathbf{k}\cdot\boldsymbol{\tau}_2} + b_3(1-i\lambda)e^{-i\mathbf{k}\cdot\boldsymbol{\tau}_3}$, $M_{AC}(\mathbf{k}) = b_1(1+i\lambda)e^{i\mathbf{k}\cdot\boldsymbol{\tau}_3} + b_4(1-i\lambda)e^{-i\mathbf{k}\cdot\boldsymbol{\tau}_2}$,

$M_{A'B}(\mathbf{k}) = b_2(1+i\lambda)e^{-i\mathbf{k}\cdot\tau_2} + b_3(1-i\lambda)e^{i\mathbf{k}\cdot\tau_3}$, and $M_{AC}(\mathbf{k}) = b_1(1+i\lambda)e^{-i\mathbf{k}\cdot\tau_3} + b_4(1-i\lambda)e^{i\mathbf{k}\cdot\tau_2}$.

We assume an orthorhombic crystal structure with lattice constants $b$, $a$, and $c$ in the $x$, $y$, and $z$ directions, respectively. The lattice displacement vectors are defined as $\tau_1 = (0, a/2, 0)$, $\tau_2 = (b/2, -a/4, 0)$, $\tau_3 = (b/2, a/4, 0)$, and $\mathbf{c} = (0, 0, c)$. The site energy is assumed to be zero. The in-plane transfer integrals are chosen as $a_1 = -0.038$ eV, $a_2 = +0.080$ eV, $a_3 = -0.018$ eV, $b_1 = +0.123$ eV, $b_2 = +0.146$ eV, $b_3 = -0.070$ eV, and $b_4 = -0.025$ eV [1], so as to reproduce a 2D DF in $\alpha$-(ET)$_2$I$_3$ when $t_A = t_{A'} = t_B = t_C = 0$ and $\lambda = \lambda' = 0$. We usually assume uniform interlayer transfers $t_A = t_{A'} = t_B = t_C = t_0$ (= 10meV). Note that the Hamiltonian preserves TRS and SIS.

Although we have employed the parameters for the 2D DF state of $\alpha$-(ET)$_2$I$_3$ under pressure, the qualitative topological properties discussed here are not sensitive to the small parameter change as long as the symmetry and band configuration are not altered. Therefore, we also use the same parameters to discuss $\alpha$-(BETS)$_2$I$_3$ in order to extract the effect of in-plane SOC. Furthermore, in our calculations we assume rather large values for the interlayer transfer integral $t_0 = 10$ meV, the in-plane SOC strength $\lambda = 0.2$, and the interlayer SOC strength $\lambda' = 1.0$ in order to enlarge the effects and make them more visible. The qualitative topological properties are not affected in this case either.

First, we study the effect of interlayer coupling in $\alpha$-(ET)$_2$I$_3$, where the in-plane SOC is negligible ($\lambda = 0$). When the interlayer SOC is also negligible ($\lambda' = 0$), the dispersion of the valence band (the third band $E_3(\mathbf{k})$) and the conduction band (the fourth band $E_4(\mathbf{k})$) is shown in Fig. 1(b) and (c). These bands have twofold spin degeneracy. The in-plane $k_x$-$k_y$ dispersion shows 2D massless DF behavior at any $k_z$, and the Dirac point does not depend on $k_z$. Thus, two straight nodal lines parallel to the $k_z$-axis are formed in the 3D Brillouin zone (BZ). No Berry curvature appears in these bands except for the nodal lines. In the case that the interlayer transfers are not uniform under SIS, i.e. $t_A = t_{A'}$, $t_B$, and $t_C$ are different, the 2D Dirac points in the $k_x$-$k_y$ plane still exist for any $k_z$, but depend on $k_z$. As a result, the nodal lines in the 3D BZ are no longer straight but curved.

Once the interlayer SOC becomes finite ($\lambda' \neq 0$), a gap opens along each nodal line as shown in Fig. 2(a) and (b). It leaves two nodal points at $ck_z = 0$ and $\pm\pi$ where the spin mixing $\Gamma(\mathbf{k})$ due to the interlayer SOC vanishes. Two spin subbands degenerating to each band can be distinguished by the value of $\sigma_x$. The texture of Berry curvature vectors $\mathbf{\Omega}_4(\mathbf{k}, \sigma_x)$ of the $\sigma_x = +1$ and $-1$ subbands in the conduction band are illustrated in Fig. 3(a) and (b). The Berry curvatures point in opposite directions in the $\sigma_x = +1$ and $-1$ subbands. The nodal point becomes a source or sink of the Berry curvature flow depending on its chirality. In the conduction band, the sink (source) of Berry curvature flow corresponds to right-handed (left-handed) chirality. Thus, the $\sigma_x = \pm 1$ subbands are found to have different chirality at the nodal point. The total Berry curvature around the nodal point is cancelled out in the degenerate band. This is the typical feature of 3D Dirac semimetals [22, 23]. Therefore, $\alpha$-(ET)$_2$I$_3$ with the interlayer coupling accompanied by interlayer SOC becomes a 3D Dirac semimetal under TRS and SIS. It can exhibit the observed chiral-anomaly-related transport phenomena such as the longitudinal negative magnetoresistance or planar Hall effect.

It should be noted that the present model approximates the orthorhombic crystal structure rather than the triclinic structure of the real $\alpha$-type crystal [2]. In general, the Dirac semimetal only appears under the TRS and SIS, but an additional symmetry is required for the Dirac point to be robust [22, 23]. In the orthorhombic model, the rotational symmetry protects the Dirac point from the hybridization of different chiralities. Therefore, in the real crystal, a small gap may open at the Dirac points.

We have discussed above the Dirac semimetal state in $\alpha$-(ET)$_2$I$_3$ under the TRS and SIS. When the SIS is broken in the interlayer transfers, which is represented by $t_A = t_0(1 + \delta)$ and $t_{A'} = t_0(1 - \delta)$ with $\delta \neq 0$, another type of the nodal-point semimetal appears. In our model, the nodal points satisfy $\delta \cos ck_z + \sigma_x \lambda' \sin ck_z = 0$. In the case of $\delta \neq 0$, $\alpha$-(ET)$_2$I$_3$ becomes a Weyl semimetal, in which the spin degeneracy is removed and the number of nodal points is doubled, as shown in Fig. 4(a) and (b). In particular, when the system has no interlayer SOC ($\lambda' = 0$), the $\sigma_x = \pm 1$ subbands degenerate with the same Berry curvature. This leads to the spin-degenerate Weyl point with identical chirality, as shown in Fig. 4(c) and (d). Note that this is not a Dirac semimetal since the total Berry curvature around the nodal point is finite. This implies that by introducing SIS breaking in the interlayer transfers, the Weyl semimetal can be realized in $\alpha$-(ET)$_2$I$_3$ without SOC. The chiral-anomaly-related phenomena are also expected in this case. The SIS breaking could be caused by the electron correlation effect.

Next, we study the effect of interlayer coupling in $\alpha$-(BETS)$_2$I$_3$, where the in-plane SOC is not negligible ($\lambda \neq 0$) and a topological SOC gap opens at the 2D Dirac points. As mentioned above, we use the same parameters for $\alpha$-(ET)$_2$I$_3$ for comparison since they do not change the topological properties of $\alpha$-(BETS)$_2$I$_3$. When the interlayer SOC is negligible ($\lambda' = 0$), the system can simply be considered as a stack of 2D TI layers. As long as the interlayer dispersion width ($\sim 4t_0$) is smaller than the SOC gap, the system is a 3D weak TI with helical surface states only on the side surfaces. A constant SOC gap opens along the nodal lines in the case of $\lambda = \lambda' = 0$ (Fig. 1(c)), and the Berry curvature vectors have no $k_z$ component. When finite interlayer SOC is introduced ($\lambda' \neq 0$), the gap depends on $k_z$, and the Berry curvature has finite $k_z$ component. However, it is not clear whether this is a weak or strong TI. According to the previous first-principles calculation, it has been discussed that it is a weak TI [8].

Since the system has TRS and SIS, we can use the parity method of Fu and Kane to check the possibility of the strong TI [24]. We focus on the parity $P_n(\mathbf{k}_{TRIM})$, which is the eigenvalue (+1 or −1) of the inversion operator, at eight time-reversal invariant wave numbers (TRIMs) $\mathbf{k}_{TRIM}$ for the $n$-th spin-degenerated band. In general, 3D TIs are characterized by a set of $Z_2$ invariants ($v_0$; $v_x$, $v_y$, $v_z$) [25]. The main invariant $v_0$ is given by $(-1)^{v_0} = \Pi P_n(\mathbf{k}_{TRIM})$, where the product is taken for all of eight TRIMs (shown in Fig. 3) and all of occupied bands ($n$ = 1, 2, 3). The subsidiary invariants $v_x$, $v_y$, and $v_z$ are also obtained from the similar formula, but the product is taken for four TRIMs at the BZ boundary. If $v_0$ = 1, the system is a 3D strong TI. If $v_0$ = 0 but there are non-zero $v_x$, $v_y$, or $v_z$, the system is a 3D weak TI.

We have applied this method to the present model of $\alpha$-(BETS)$_2$I$_3$ with $\lambda' \neq 0$ using the inversion operator [26], resulting in the $Z_2$ invariants ($v_0$; $v_x$, $v_y$, $v_z$) = (0; 0, 0, 1). This implies that $\alpha$-(BETS)$_2$I$_3$ remains a 3D weak TI under interlayer SOC as long as TRS and SIS are preserved as the preceding work [8]. Therefore, the experimentally suggested 3D strong TI state with surface state surrounding entire surfaces seems to be hardly realized under TRS and SIS. The topological transition accompanied by the symmetry breaking, possibly due to electron correlation, seems necessary to explain the strong TI state in $\alpha$-(BETS)$_2$I$_3$. It may correspond to the anomalous structure in the temperature dependence of the resistance observed in $\alpha$-(BETS)$_2$I$_3$ [15].

In conclusion, we have studied a 3D tight-binding model for $\alpha$-type organic conductors $\alpha$-(ET)$_2$I$_3$ and $\alpha$-(BETS)$_2$I$_3$, considering interlayer coupling (hopping) accompanied by SOC. The interlayer SOC originates from the I$_3^-$ anion potential. This model preserves TRS and SIS. In $\alpha$-(ET)$_2$I$_3$, which has no in-plane SOC, the system becomes a 3D nodal-line semimetal if it also has no interlayer SOC. Once the interlayer SOC becomes finite, it becomes a 3D Dirac semimetal with nodal points of degenerate chirality. In contrast, it is found that SIS breaking in interlayer transfers realizes the 3D Weyl semimetal even without SOC. In $\alpha$-(BETS)$_2$I$_3$, which has finite in-plane SOC, the system becomes a 3D weak TI regardless of the interlayer SOC. It seems difficult to explain the proposed 3D strong TI assuming TRS and SIS. The 3D strong TI state in $\alpha$-(BETS)$_2$I$_3$ could be realized by some topological transition accompanied by the symmetry breaking which is possibly caused by electron correlation effect.

**Acknowledgements**

The author thanks Dr. M. Kashiwagi for his valuable comment. This work was partially supported by JSPS KAKENHI Grant Numbers JP23K03297 and JP24H01610.

# References

*osada@issp.u-tokyo.ac.jp

[1] For a review see, K. Kajita, Y. Nishio, N. Tajima, Y. Suzumura, and A. Kobayashi, J. Phys. Soc. Jpn. **83**, 072002 (2014).

[2] K. Bender, I. Hennig, D. Schweitzer, K. Dietz, H. Endres, and H. J. Keller, Mol. Cryst. Liq. Cryst. **108**, 359 (1984).

[3] S. Katayama, A. Kobayashi, and Y. Suzumura, J. Phys. Soc. Jpn. **75**, 054705 (2006).

[4] A. Kobayashi, S. Katayama, Y. Suzumura, and H. Fukuyama, J. Phys. Soc. Jpn. **76**, 034711 (2007).

[5] T. Mori, J. Phys. Soc. Jpn. **79**, 014703 (2010).

[6] T. Kawarabayashi, Y. Hatsugai, T. Morimoto, and H. Aoki, Phys. Rev. B **83**, 153414 (2011).

[7] H. Kino and T. Miyazaki, J. Phys. Soc. Jpn. 75, 034704 (2006).

[8] S. Kitou, T. Tsumuraya, H. Sawahata, F. Ishii, K. Hiraki, T. Nakamura, N. Katayama, and H. Sawa, Phys. Rev. B **103**, 035135 (2021).

[9] T. Tsumuraya and Y. Suzumura, Eur. Phys. J. B **94**, 17 (2021).

[10] N. Tajima, T. Yamauchi, T. Yamaguchi, M. Suda, Y. Kawasugi, H. M. Yamamoto, R. Kato, Y. Nishio, and K. Kajita, Phys. Rev. B **88**, 075315 (2013).

[11] Y. Kawasugi, H. Masuda, M. Uebe, H. M. Yamamoto, R. Kato, Y. Nishio, and N. Tajima, Phys. Rev. B **103**, 205140 (2021).

[12] S. M. Winter, K. Riedl, and R. Valentí, Phys. Rev. B **95**, 060404 (2017).

[13] T. Osada, J. Phys. Soc. Jpn. **87**, 075002 (2018).

[14] C. L. Kane and E. J. Mele, Phys. Rev. Lett. **95**, 226801 (2005).

[15] T. Nomoto, S. Imajo, H. Akutsu, Y. Nakazawa, and Y. Kohama, Nat. Comms. **14**, 2130 (2023).

[16] T. Morinari, J. Phys. Soc. Jpn. **89**, 073705 (2020).

[17] Y. Nakamura and T. Morinari, J. Phys. Soc. Jpn. **92**, 123701 (2023).

[18] S. Raghu, X. L. Qi, C. Honerkamp, and S. C. Zhang, Phys. Rev. Lett. **100**, 156401 (2008).

[19] D. Ohki, K. Yoshimi, and A. Kobayashi, Phys. Rev. B **105**, 205123 (2022).

[20] N. Tajima, Y. Kawasugi, T. Morinari, R. Oka, T. Naito, and R. Kato, J. Phys. Soc. Jpn. **92**, 013702 (2023).

[21] N. Tajima, Y. Kawasugi, T. Morinari, R. Oka, T. Naito, and R. Kato, J. Phys. Soc. Jpn. **92**, 123702 (2023).

[22] S. Murakami, New J. Phys. **9**, 356 (2007).

[23] B.-J. Yang and N. Nagaosa, Nat. Comms. **5**, 4898 (2014).

[24] L. Fu and C. L. Kane, Phys. Rev. B **76**, 045302 (2007).

[25] L. Fu, C. L. Kane, and E. J. Mele, Phys. Rev. Lett. **98**, 106803 (2007).

[26] F. Piechon and Y. Suzumura, J. Phys. Soc. Jpn. **82**, 033703 (2013).

**Figure 1** (Osada)

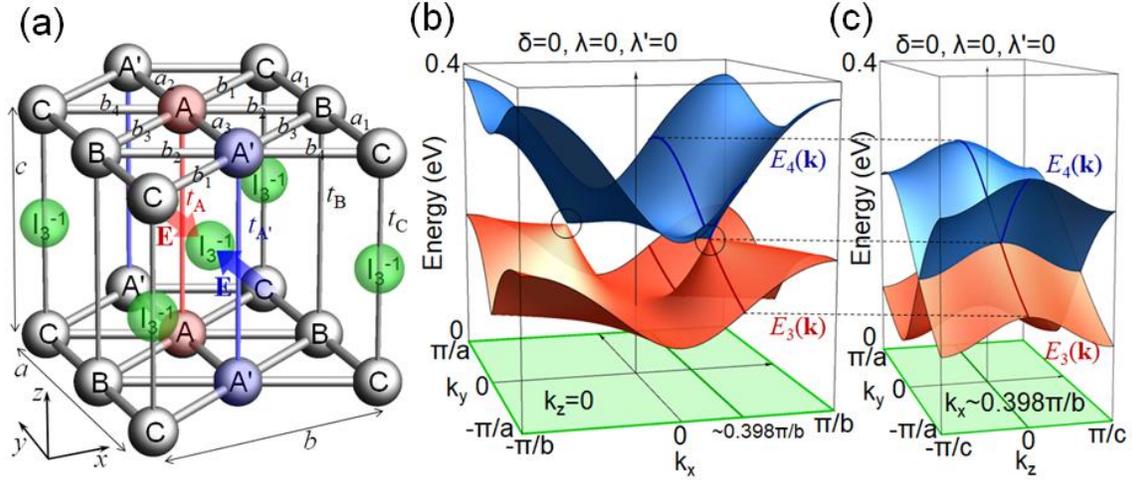

**FIG. 1.** (color online)

(a) Schematic of the 3D lattice structure of ET or BETS molecules in α-(ET)$_2$I$_3$ or α-(BETS)$_2$I$_3$. In-plane and interlayer transfer integrals are indicated. The I$_3^-$ anions and the electric field **E** on the A-A and A'-A' hopping paths are also shown. (b)(c) Dispersion of the valence and conduction bands in α-(ET)$_2$I$_3$ ($\lambda = 0$) in the absence of interlayer SOC ($\lambda' = 0$). Each band has twofold spin degeneracy. (b) In-plane ($k_x$-$k_y$) dispersion at $k_z = 0$. Solid circles denote Dirac points. (c) Interlayer ($k_z$-$k_y$) dispersion at the $k_x$ value ($k_x \sim 0.398\pi/b$) of an in-plane Dirac point in (b). Dirac points are present at all $k_z$ and form straight nodal lines parallel to the $k_z$ axis in the 3D BZ.

**Figure 2** (Osada)

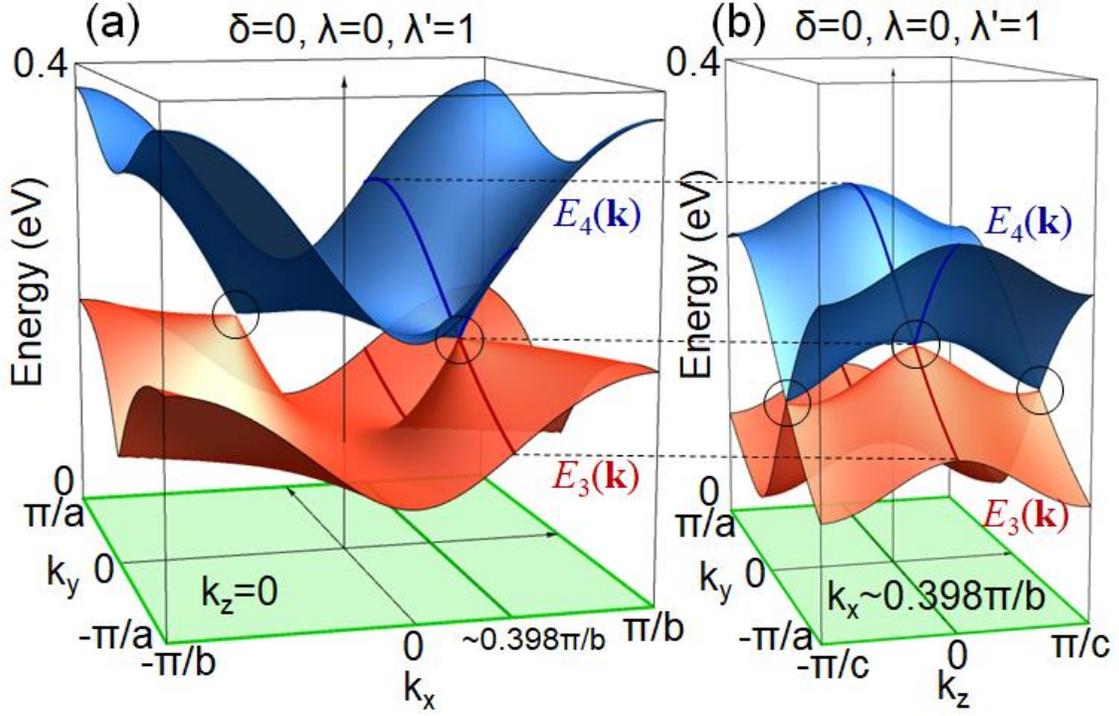

**FIG. 2.** (color online)

Dispersion of the valence and conduction bands in α-(ET)$_2$I$_3$ ($\lambda = 0$) in the case of finite interlayer SOC ($\lambda' = 1.0$). Each band has twofold spin degeneracy. (a) In-plane ($k_x$-$k_y$) dispersion at $k_z = 0$. Solid circles denote Dirac points. (b) Interlayer ($k_z$-$k_y$) dispersion at $k_x \sim 0.398\pi/b$. Dirac points exist at $k_z = 0$ and $\pm\pi/c$ as indicated by solid circles.

**Figure 3** (Osada)

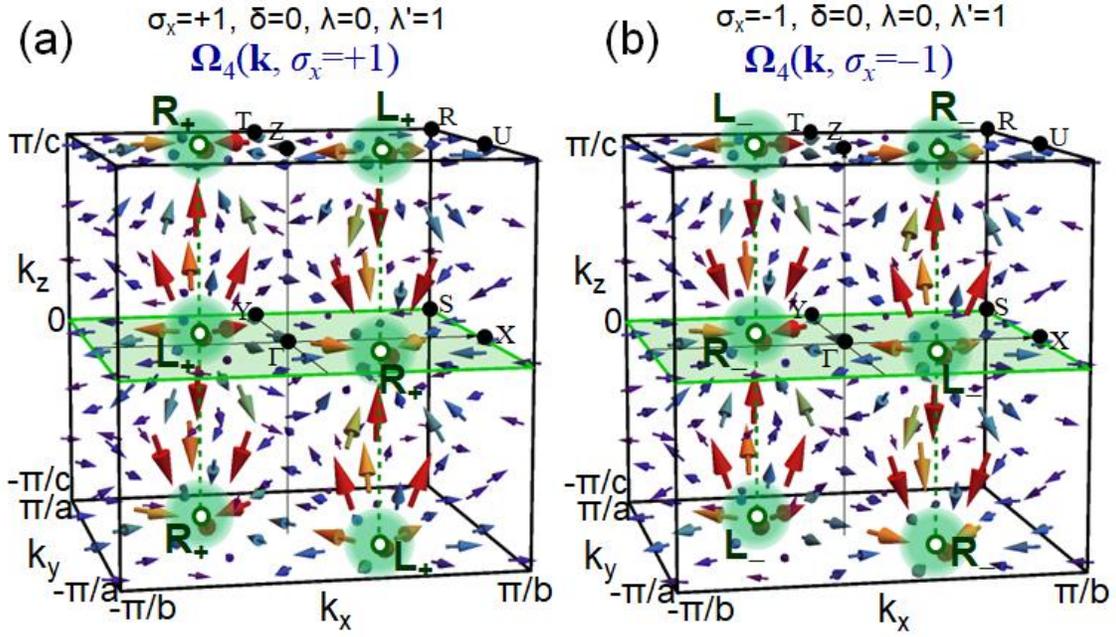

**FIG. 3.** (color online)

Calculated Berry curvature texture of (a) $\sigma_x$=+1 spin subband and (b) $\sigma_x$=−1 spin subband of the conduction band in $\alpha$-(ET)$_2$I$_3$ ($\lambda$ = 0) with finite interlayer SOC ($\lambda'$ = 1.0). The chirality of the nodal points is indicated by "R" or "L" with the subscript "±" indicating $\sigma_x$. The nodal points in (a) and (b) have the same positions with opposite chirality, forming Dirac points. Eight TRIMs are also shown.

**Figure 4** (Osada)

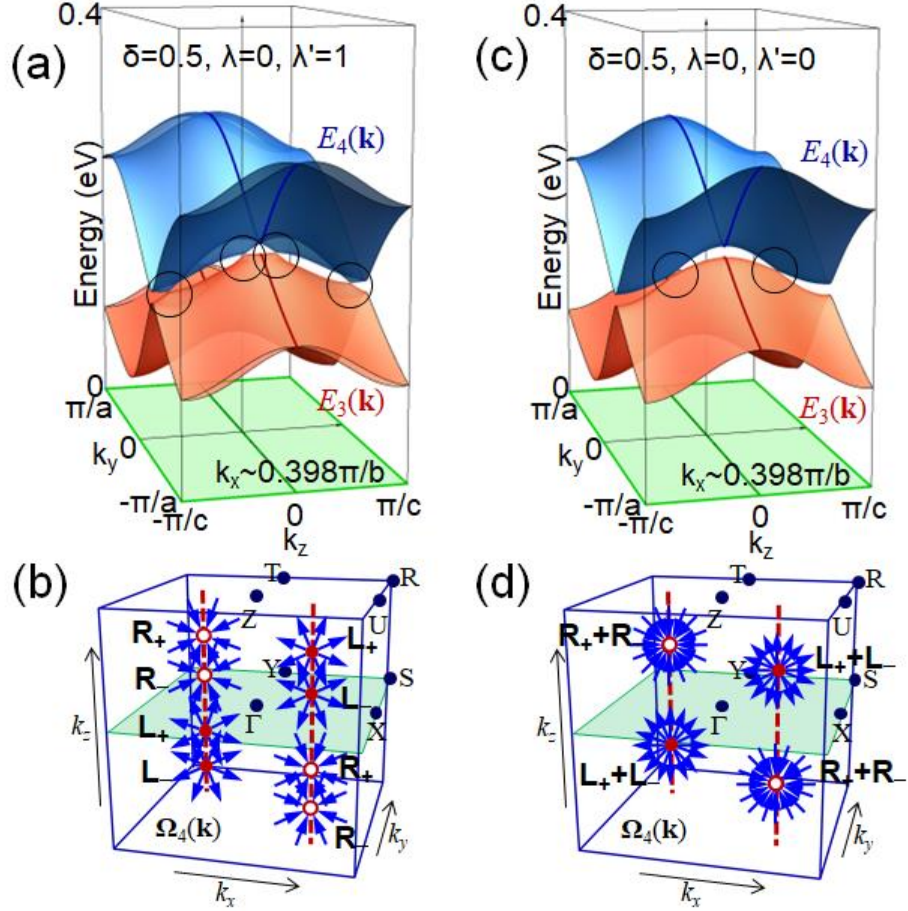

**FIG. 4.** (color online)

(a)(c) Interlayer dispersion of the valence and conduction bands and (b)(d) schematic Berry curvature around the Weyl points in the 3D BZ when the interlayer transfer breaks SIS ($\delta$ = 0.5) in $\alpha$-(ET)$_2$I$_3$ ($\lambda$ = 0). (a) and (b) show the case of finite interlayer SOC ($\lambda$' = 1.0), and (c) and (d) shows the case of no interlayer SOC ($\lambda$' = 0). In these panels, the $\sigma_x$=+1 and $\sigma_x$=−1 spin subbands are shown superimposed. In (a) and (c), the Weyl points are indicated by solid circles. In (b) and (d), the Weyl points of each spin subband ($\sigma_x$=±1) are labelled "R$_\pm$" or "L$_\pm$" depending on their chirality (right-handed or left-handed). Red dashed lines indicate the original nodal line.